\documentclass[aps,prl,preprint,groupedaddress]{revtex4}
\usepackage{epsfig}
\usepackage{latexsym}
\usepackage{amssymb}
\begin{document} 

\title{Support-mediated activation and deactivation of Pt thin films} 

\author{Valentino R. Cooper}
\author{Alexie M. Kolpak}
\author{Yashar Yourdshahyan}
\author{Andrew M. Rappe}
\email{rappe@sas.upenn.edu}
\affiliation{The Makineni Theoretical Laboratories, Department of Chemistry, University of Pennsylvania,\\231 S. 34th Street, Philadelphia, PA 19104-6323}

\date{\today}

\begin{abstract}
Using ab initio methods, we examine the the charge distribution at the
interface of alpha-alumina-supported Pt films, and we consider the
influence of this interface on CO adsorption.  We demonstrate that a
combination of electrostatic charge transfer and covalent bonding
governs the interfacial interactions, and that these interactions play
an important role in the metal reactivity.  By modifying the interface
and varying the Pt film thickness over a nanoscale range, CO
adsorption can be significantly enhanced or diminished.  These
observations could be used to tune the reactivity of Pt particles.
\end{abstract}

\maketitle

\section{Introduction}
The ability to control the interactions of molecules on metal surfaces
provides new opportunities for the advance of many technological and
industrial processes.  Such knowledge will allow for the creation of
more efficient catalysts for manufacturing processes, fuel combustion,
and the treatment of automotive exhausts and for the production of
more sensitive chemical sensors.  As a result, the prototypical
reaction of CO adsorbing to transition metal surfaces has received
much attention.  Blyholder used H\"{u}ckel molecular orbital theory to
describe the mechanism by which CO molecules bind to a metal
surface~\cite{Blyholder64p2772}, thus explaining shifts in CO
stretching frequencies for CO binding to various sites.  Here, CO
binding to the top site on a metal surface involves both the donation
of electrons from the filled CO 5$\sigma$ molecular orbital to the
metal $d_{z^2}$ orbitals (direct bonding), and the back-donation of
electrons from the metal $d_{xz}$ and $d_{yz}$ orbitals into the CO
2$\pi^*$ anti-bonding orbitals.  CO binding at metal hollow sites is
dominated by back-bonding from the metal $d_{xz}$ and $d_{yz}$
orbitals, with a small contribution from direct bonding to the metal
$d_{xy}$ and $d_{x^2-y^2}$ orbitals.

Hammer and N{\o}rskov (HN) later extended this work to illustrate the
relationship between the strength of the CO-metal bond and the
electronic properties of the metal~\cite{Hammer95p211, Hammer96p2141}.
Using density functional theory (DFT)\cite{Hohenberg64p864,
Kohn65pA1133}, they demonstrated that the binding energy of molecules
to transition metal surfaces can be correlated to shifts in the
centers of the metal $d$-bands.  Since this property could be measured
using spectroscopic techniques, their model allowed for the prediction
of CO binding strengths using simple surface experiments.

Nanoparticles and thin films further broaden the study of
molecule-metal interactions, as nano-dimensional metal particles often
exhibit distinctly different properties than bulk.  Investigations of
gold and platinum nanoparticles indicate that their reactivity is
strongly dependent on their size and geometry~\cite{Mills02p493,
Wang97p302, Yoo03p1, Yoon03p4066, Haruta97p153, Valden98p1647,
Sanchez99p9573}.  In many cases, these particles have enhanced
reactivity with decreasing size.  Numerous theoretical and
experimental studies also suggest that the support material plays an
important role in activating or deactivating the surface of metal
nanoparticles~\cite{Bozo01p393,Walter01p44, Molina03p206102-1,
Lindsay03pL859, Haruta97p153, Putna97pL1178}.  Strain studies of metal
thin films give evidence that in-plane strain alters the width of the
metal $d$-band and shifts the $d$-band center toward or away from the
Fermi level~\cite{Mavrikakis98p2819}.  In accordance with the HN
model, these shifts affect adsorption at the metal surface. Although
these studies present clear deviations in the nanosystem as a result
of changing the supporting material or the size or shape of the metal,
it is still not clear how the charge boundary at the support-metal
interface gives rise to the newly observed properties.

Recent experimental studies by Chen and Goodman of one and two atomic
layers of gold supported on TiO$_2$ show that charge transfer between
the support and the metal catalyst have a significant effect on the
reactivity at the metal surface~\cite{Chen04p252}.  Their thin film
geometry eliminated particle shape and support effects to show that
the film thickness is a crucial parameter in altering the reactivity
at the metal surface.

In this letter, we present DFT calculations to systematically
investigate the extent to which the charge distribution at the surface
of an $\alpha$-alumina support alters the electronic properties of Pt
thin films.  The Pt thin film geometry used in this study excludes
particle shape and direct support effects and allows us to concentrate
on the effect of the metal-support charge transfer on the metal
surface reactivity.  We observe dramatic differences in the electronic
properties at the metal surface in Pt films of 1-5 atomic layers
deposited on oxide supports with different electronegativities, which
emphasizes the importance of the underlying substrate.  In particular,
we show that an electropositive oxide support strengthens the bonding
of CO to the metal film, while the electronegative support exhibits
non-monotonic bonding to the thin film surface, as a function of layer
thickness, resulting in a switch in site preference between a
monolayer and a bilayer of Pt.  Furthermore, our results demonstrate
that the formation of hybrid orbitals at the metal-oxide interface is
critical in determining the adsorption properties at the metal
surface.  We also show that by explicitly considering the symmetry of
the metal $d$-states, particularly of the newly formed hybrid states,
the HN model can be extended to describe complex, strongly bound
metal-support systems.

\begin{figure}[t]
\includegraphics[height=2.2in]{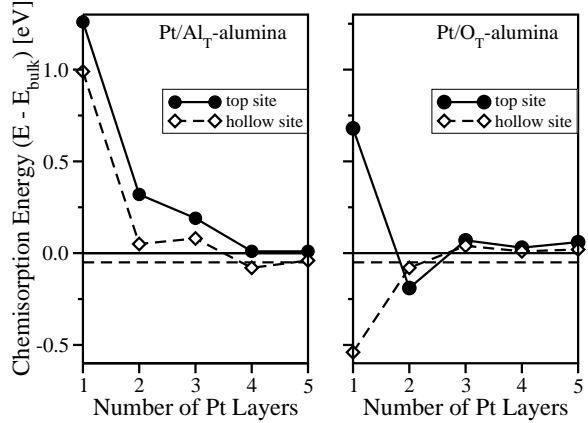}
\caption{\label{ChemE} CO chemisorption energies for the top
($\bullet$--$\bullet$) and hollow ($\diamond$- -$\diamond$) sites on
the $n$Pt/Al$_T$-Al$_2$O$_3$ (left) and $n$Pt/O$_T$-Al$_2$O$_3$
(right) systems.  Binding energies are relative to the top site of the
Pt(111) surface (---).  The (- -) line represents the binding energy
of CO at the Pt(111) hollow site.}
\end{figure}

\section{Methodology}
We examine a range of metal-oxide interface properties by considering
two variations of the $\alpha$-Alumina surface: the Al-terminated
surface (Al$_T$), which is the most stable clean
surface~\cite{Toofan98p162}, and the strongly polar O-terminated
surface (O$_T$), which is stable when hydrogen or a supported metal is
adsorbed~\cite{Eng00p1029}.  We use a slab geometry with an in-plane
$(\sqrt{3}\times\sqrt{3})R30^{\circ}$ unit cell and periodic boundary
conditions to model the alumina surface\cite{Pauling25p781}.  The
layer stacking can be represented by the formula
(Al-O$_3$-Al)$_4$-Al-O$_3$-Al$_m$-(Pt$_3$)$_n$, where $m$ = 0 or 1 for
the O$_T$ and Al$_T$ surfaces, respectively, and $n$ is the number of
Pt layers.  For both terminations, the interfacial Pt atoms are
directly above the surface O atoms, forming a Pt(111) film.  All
calculations were performed using DFT with the generalized gradient
approximation~\cite{Perdew92p6671} for the exchange correlation
functional as implemented in the dacapo code~\cite{Hammer99p7413},
with a planewave cutoff of 30 Ry and a 2$\times$2$\times$1
Monkhorst-Pack $k$-point mesh. The theoretical $\alpha$-Al$_2$O$_3$
in-plane lattice constant of 4.798~\AA\ was used (experimental =
4.759~\AA~\cite{Lee85p247}).  In order to eliminate strain effects,
all adsorption energies were compared to unsupported Pt(111)
calculations at the same in-plane geometry as the supported metal.
(This corresponds to the experimental Pt(111) in-plane lattice
constant of 2.77~\AA) For our simulations we fixed the ions in the
bottom two alumina layers to their theoretical positions, relaxed the
third layer perpendicular to the surface, and fully relaxed the
remaining alumina layers, Pt layers and adsorbates until the remaining
force on each atom was less than 0.01 eV/\AA.  We correct for known
DFT CO chemisorption errors\cite{Feibelman01p4018, Grinberg02p2264}
using the extrapolation method of Mason and
coworkers~\cite{Mason04p161401}.

\section{Results}

Figure~\ref{ChemE} shows the top site chemisorption energy for the
adsorption of CO onto the Pt/$\alpha$-alumina system as a function of
metal layers, relative to Pt(111), on the O$_T$- and
Al$_T$-$\alpha$-alumina surfaces.  For both terminations, the first
layer of Pt ions are deposited above alumina surface oxygen ions.  In
both cases, a monolayer of metal on the surface results in an
enhancement of the CO top site binding energy relative to Pt(111).
For Pt on the Al$_T$ surface, a monotonic decrease in the
chemisorption energy is observed as the Pt film thickness is
increased.  The system returns to Pt(111) behavior at four layers of
Pt.  In contrast, the adsorption of CO on the Pt/O$_T$ system is
non-monotonic.  In this case, there is a dramatic decrease in the top
site chemisorption energy and an increase in the hollow site
chemisorption energy for two layers of Pt, such that the binding of CO
to the surface is weaker than for Pt(111).  For three layers of Pt on
the O$_T$ surface, the chemisorption energy oscillates above the
Pt(111) energy, eventually returning to the Pt(111) value for $n >$ 4.

Table~\ref{site} tabulates the site preference energy ($E_{\rm{top}}$
- $E_{\rm{fcc}}$) for CO adsorption on the Pt/alumina system as a
function of metal layers.  Here we see that both the Pt/Al$_T$- and
the Pt/O$_T$-alumina monolayer systems show a stronger preference for
top site binding than the bulk material.  However, the Pt/O$_T$ system
shows a much larger preference toward top site binding than the
Pt/Al$_T$ system.  Furthermore, the Pt/Al$_T$ system binding energies
asymptote to that of Pt(111) with increasing film thickness, while
the Pt/O$_T$ bilayer exhibits a switch in site preference.
Table~\ref{site} also indicates a much greater effect on the site
preference energy, as thicker films ($n >$ 5) are needed before the
Pt/O$_T$ system fully returns to Pt(111).

\begin{table}[b]
\begin{tabular}{c|c|c|ccc|ccc}
\hline
No. Pt & Al$_T$ Site &O$_T$ Site &Al$_T$&Term.&(eV)&O$_T$&Term.&(eV)\\Layers &  Pref.&   Pref.&   $d_{z^2}$    & $d_{xy}$& $d_{xz}$ &  $d_{z^2}$ & $d_{xy}$& $d_{xz}$ \\
\hline
\hline
1 & 0.27 & 1.22 & 0.49 & 0.54 &  0.55 &  0.36 & -0.23 & -0.78\\
\hline
2 & 0.27 & -0.11 & -0.04 & 0.09 &  0.05 &  0.14 &  0.00 & -0.07\\
\hline
3 & 0.11 & 0.03 &-0.10 & 0.02 &  0.03 &-0.06 &  0.01 &  0.03\\
\hline
4 & 0.09 & 0.02 & 0.01 & 0.00 & -0.01 &  0.03 &  0.01 &  0.03\\
\hline
5 & 0.05 & 0.04 & 0.00 & 0.00 & 0.00 &  0.00 &  0.00 & -0.01\\
\hline
\hline 
Pt(111)& 0.05 & 0.05 & -1.35 & -2.03 & -1.64 & -1.35 & -2.03 & -1.64\\
\hline
\end{tabular}
\caption{CO site preference energy, $E_{\rm{top}}$-$E_{\rm{fcc}}$, and shifts in the $d-$ band centers relative to Pt(111) for the $n$Pt/Al$_2$O$_3$ system.  All energies are in eV.}
\label{site}
\end{table}

\section{Discussion}

When a transition metal is deposited onto a support material, the
metal $d$-states change as the metal interacts with the substrate.
According to the HN model, upward or downward shifts in the metal
$d$-band center can cause corresponding increases or decreases in the
bonding interaction of molecules with the metal surface. To apply the
HN model to describe the Pt/Al$_2$O$_3$ systems, we decompose the
metal $d$-band and consider the shifts of the individual $d$-orbitals.
These are summarized in Table 1.

\begin{figure}
\includegraphics[height=2.50in]{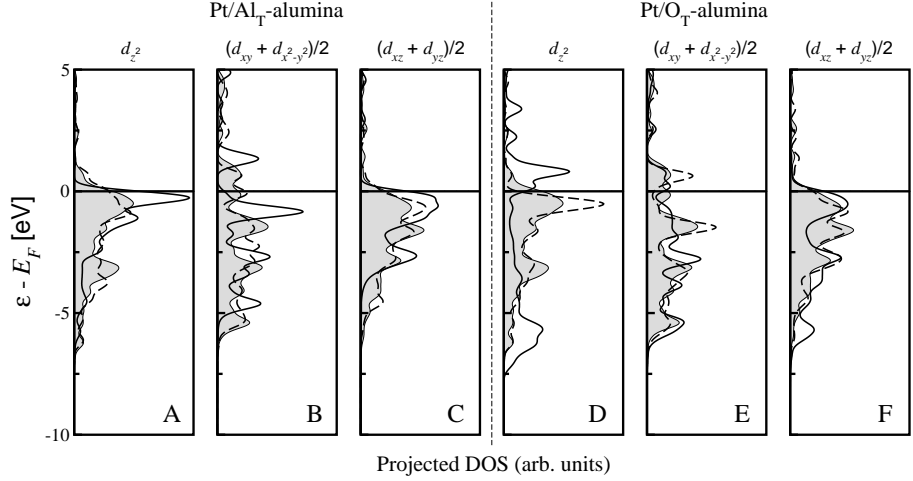}
\caption{\label{DOS} The density of states (DOS) plots for the surface
Pt atoms for the Pt/$\alpha$-alumina system. (A-C) are projected DOS
for the Pt/Al$_T$ system and (D-F) are projected DOS for the Pt/O$_T$
system.  (A) and (D) depict the DOS projected onto the $d_{z^2}$
orbitals, (B) and (E) are the average of the DOS projected onto the
$d_{xy}$ and $d_{x^2-y^2}$ orbitals, and (C) and (F) are the average
of the DOS projected onto the $d_{xz}$ and $d_{yz}$ orbitals for each
system.  The shaded regions represent the Pt(111) surface, the solid
(---) lines are for the monolayer and the dashed (-~-) lines are for
the bilayer.}
\end{figure}

When a single layer of Pt is deposited on the Al$_T$ surface, the
electropositive Al ions donate charge to the surface metal atoms.
This transfer of electrons results in an upward shift in the metal
$d$-band center, relative to Pt(111).  We can decompose the $d$-bands
of the metal surface atoms to consider the orbitals involved in CO
bonding.  In Table~\ref{site} we see that there is a large upward
shift (0.55 eV) in the $d_{xz}$ and $d_{yz}$ band centers relative to
Pt(111), as can be seen in the large increase in the area under the
peak just below the Fermi level.  This shift makes more states of
appropriate energy available for back donation to the CO 2$\pi^*$
orbital, greatly enhancing the binding at the top and hollow sites.
While such a large increase in back-bonding might suggest a more
pronounced enhancement of hollow site binding energy (and thus a
decrease in the site preference energy), this is not the case because
of the nature of the direct bonding at the top and hollow sites.  The
number of free states in the $d_{z^2}$ orbitals (Figure~\ref{DOS}A) is
similar to that in Pt(111), resulting in negligible changes in the
direct bonding contribution to the top site binding energy.  On the
other hand, interaction of the $d_{xy}$ and $d_{x^2-y^2}$ orbitals
with the surface Al atom results in a large upwards shift in these
states, reducing the number of free states near the Fermi level
accessible for receiving electrons from the CO 5$\sigma$
(Figure~\ref{DOS}B). This loss in free states, combined with the small
spatial overlap between these orbitals and the CO 5$\sigma$,
significantly decreases the direct-bonding contribution at the hollow
site.  Consequently, there is a much larger top site preference for
this system compared to unsupported Pt(111) (Figure~\ref{site}).

The bilayer of Pt metal shows a shift of all of the metal $d$-orbitals
(Figure~\ref{DOS}A-C) back towards their Pt(111) arrangement.  The
$d_{xz}$ and $d_{yz}$ are still shifted up by 0.05 eV relative to
Pt(111) (Table~\ref{site}), exhibiting stronger back-bonding.
Furthermore, there are more free $d_{z^2}$ states than Pt(111)
available for direct-bonding, while there is a decrease in the number
of free $d_{xy}$ and $d_{x^2-y^2}$ orbitals relative to Pt(111).  This
again causes there to be a stronger preference towards top site
binding (Figure~\ref{site}).  For three or more layers of Pt, the DOS
of the surface atoms returns to the Pt(111) value, as do the top and
hollow site chemisorption energies.

Figure~\ref{ChemE} shows that for a single layer of Pt on the O$_T$
surface there is an increase in CO top site binding strength by 0.69
eV relative to Pt(111).  However, an analysis of the DOS for the one
layer Pt/O$_T$ system shows that the $d$-band center shifts down by
0.60 eV relative to Pt(111).  These results can not be described by
the basic HN model bonding picture, which would predict a decrease in
the chemisorption energy.  To understand the origin of this
difference, we need to consider the bonding at the metal-oxide
interface and its effect on the surface metal $d$-states.  

When a platinum atom is deposited onto the O$_T$-alumina surface,
there is a transfer of electrons from the platinum atoms to the more
electronegative oxygen atoms.  More importantly, this interaction
results in the formation of hybrid orbitals between the Pt metal and
the oxide surface atoms.  Figures~\ref{DOS} D-F show the projected DOS
for the top layer Pt metal $d$-orbitals as a function of layer
thickness.  Here we see, for the monolayer, the creation of hybrid
orbitals causes a splitting of both the d$_{z^2}$ (Figure~\ref{DOS}D)
and the $d_{xz}$ and $d_{yz}$ (Figure~\ref{DOS}F) orbitals.  The
downshift in the $d$-band center is related to the large splitting of
the $d$-orbitals into these hybrid orbitals, creating a higher energy
anti-bonding orbital and a much lower energy bonding orbital.  The new
$d_{z^2}$ anti-bonding orbital has the correct symmetry for bonding
with the CO 5$\sigma$ orbital and has a much greater number of free
states available to accept electrons from the CO molecule.  This
strengthens the direct bonding between the CO molecule and the metal
surface and explains the large increase in the CO top site
chemisorption energy.  Hollow site bonding, on the other hand, is more
affected by changes in the remaining $d$-orbitals.  The overall
downward shift in $d_{xz}$ and $d_{yz}$ band centers
(Table~\ref{site}) reduces the amount of electrons of suitable energy
available for back-donation.  Simultaneously, there is a reduction in
the number of free $d_{xy}$ and $d_{x^2-y^2}$ orbitals, further
inhibiting the direct binding at the hollow site.  These effects are
manifested in the huge site preference for the single layer Pt/O$_T$
system.

For the bilayer of Pt on O$_T$-alumina, we see that there is still a
downward shift of the $d$-band center relative to Pt(111),
corresponding to the observed decrease in top site chemisorption
energy.  In this case, the top layer is not directly in contact with
the oxide support, and no hybrid orbitals are formed.  However, the
surface Pt atoms interact with the hybrid orbitals of the interfacial
Pt atoms, resulting in a shift of their $d_{z^2}$ orbitals below the
Fermi level (Figure~\ref{DOS}D).  The dramatic reduction in free
$d_{z^2}$ states causes a decrease in the top site binding energy
relative to Pt(111).  On the other hand, the $d_{xz}$ and $d_{yz}$
(Figure~\ref{DOS}F) orbitals now resemble Pt(111) states, restoring
most of the bulk back-bonding character.  Additional shifts in the
$d_{xy}$ and $d_{x^2-y^2}$ orbitals at the Fermi level reduces the
direct bonding at the hollow site (similar to the 1LPt/Al$_T$ case);
therefore the hollow site binding energy is slightly less than
observed for Pt(111).  The large weakening of top site binding and the
small difference in hollow site binding result in a change in the site
preference of Pt/O$_T$ bilayers.  Similar to Pt/Al$_T$, both the
chemisorption and site preference energies begin to return to Pt(111)
values for thicker films as the top layer DOS begins to resemble that
of Pt(111).  Due to the larger charge transfer at the Pt/O$_T$
interface, and the formation of hybrid orbitals within the first layer
of Pt, there is less charge screening in the Pt thin film, causing a
slower return to Pt(111) values for $n >$ 5.

\section{Conclusion}
In conclusion, we have shown how changes in the charge distribution at
the metal-support interface affect adsorption of CO on the surface of
Pt metal thin films.  Our results present a theoretical basis for
beginning to understand the interactions between an oxide support and
the properties at the metal surface.  Using a modified HN model, we
show that the changes in the metal surface properties can be
correlated to the changes in the electronic properties of the metal
$d-$states as they interact with the oxide support.  While it is known
that low-coordinated sites greatly enhance the reactivity of metal
particles~\cite{Wang97p302, Yoo03p1}, these results demonstrate the
importance of metal-support charge transfer in defining the properties
at the metal surface, offering further support for the recent work of
Chen and Goodman. In addition, our findings suggest that increased
reactivity at the perimeter of metal particles with diameters $<$ 5
nm~\cite{Haruta97p153, Valden98p1647, Heiz99p3214, Putna97pL1178} may
be partially attributed to the strong metal-oxide coupling accessible at
these boundaries.  

The very different CO adsorption behavior as a
function of Pt film thickness for the two surface terminations shows
that coupling between the metal and oxide support is also sensitive to
the surface polarization of the oxide.  The fact that these results
are greatly diminished at four or five layers indicates that this is a
nanoscale effect.  Our study demonstrates the wide range of chemical
activity that can be accessed by manipulating the oxide support and
the thickness of the metal film, offering numerous possibilities for
the design of more reactive/selective catalysts for applications as
diverse as chemical sensors, fuel cells, and photochemical reactions.
\section{Acknowledgments}
\begin{acknowledgments}
We thank Sara E. Mason and Ilya Grinberg for discussions on CO
adsorption energy corrections. This work was supported by the Air
Force Office of Scientific Research, under Grant No. FA9550-04-1-0077,
and the Office of Naval Research, under Grant No. N-000014-00-1-0372.
Computational support was provided by the High-Performance Computing
Modernization Office of the Department of Defense and the Defense
University Research Instrumentation Program. A.M.R. thanks the Camille
and Henry Dreyfus Foundation for support.  V.R.C. thanks IBM and ACS
for support through the IBM Graduate Student Award in Computational
Chemistry.  A.M.K. was supported by a GAANN fellowship.
\end{acknowledgments}

\end{document}